\documentstyle[prd,aps,preprint,psfig]{revtex}

\begin{document}

%@twocolumnfalse\endcsname
%%
%%
%\tighten
%\draft
%%

\title{Natural Mass Generation for the Sterile Neutrino }
\author{K.S. Babu$^1$ and T. Yanagida$^{2,3}$}
\address{$^1$Department of Physics, Oklahoma State University,
Stillwater, OK 74078, USA \\
$^{2}$Department of Physics, University of Tokyo, Tokyo 113-0033,
Japan \\   $^3$Research Center for the Early Universe, University of
Tokyo, Tokyo, 113-0033, Japan}
%%%

\date{\today}

\maketitle

\begin{abstract}
    We point out that there is a serious cosmological problem in the 
 supersymmetric standard model if a sterile neutrino is responsible
 for the solar neutrino oscillation, and propose a possible solution to this
 problem. We show that our solution induces naturally a mass of order 
 $10^{-4}$ eV for the sterile neutrino, which is deeply related to the
 mechanism of supersymmetry breaking. 
\end{abstract}

%]
\clearpage
%%%%%%%%%%%%%%%%%%%%%%%%%%%%%%%%%%%%%%%%%%%%%%%%%%%%%%%%%%%%%%%%%%%%%%
%%%%%%%%%%%%%%%%%%%%%%%%%%%%%%%%%%%%%%%%%%%%%%%%%%%%%%%%%%%%%%%%%%%%%%

Two flavor active--sterile neutrino oscillation seems to be disfavored
by recent Superkamiokande 
data on both atmospheric and solar neutrino experiments \cite{suzuki}. 
However, a recent global analysis of the solar neutrino data \cite{concha}
suggests that both the small angle MSW oscillation and the
quasi--vacuum oscillation (corresponding to $\Delta m^2 \simeq
10^{-7}-10^{-9}$ eV$^2$) are still consistent solutions.   
It has also been pointed out recently \cite{barger}
that an energy--independent active--sterile neutrino oscillation 
is well consistent with 
the present solar neutrino experiments 
with the exception of the $^{37}C\ell$ results. Thus, the
sterile neutrino is still interesting, since it may explain all neutrino 
oscillation data including LSND experiments\cite{LSND}.
In this short paper
we propose a natural mechanism for generating a small mass for the
sterile neutrino $\nu_{s}$, which induces a  $\nu_e$-$\nu_s$ oscillation
together with the conventional seesaw mechanism \cite{yanagida}. This
new mechanism is deeply related to the dynamics of supersymmetry (SUSY) 
breaking.

Before describing the model let us discuss a cosmological difficulty due to 
the presence of the sterile neutrino in the SUSY standard model. Since
the sterile neutrino is a gauge singlet and its Yukawa coupling constant 
is very small ($y_s \simeq 10^{-15}$)\footnote{This small Yukawa
coupling, $W=y_sLHS$, induces a small Dirac-type mass for the sterile
neutrino of order $10^{-4}$ eV, which is required for the 
solar neutrino quasi--vacuum 
oscillation. $y_s$ should be similar in magnitude for the small angle
MSW solution as well.}, the scalar partner of $\nu_s$ has 
a very flat potential. Thus, it is quite natural to consider that it has 
a large value of order the Planck scale ($M_G \simeq 2.4 \times 10^{18}$ 
GeV) at the end of inflation and its coherent oscillation dominates the 
early universe like moduli fields in string theory. The lifetime of the
scalar sterile neutrino is estimated as $\tau \simeq 10^4 $ sec with the
above small Yukawa coupling constant $y_s$ and a mass $m\simeq 1$ TeV. 
It is well known that such late decays of massive heavy particles
destroy the success of the big-bang
nucleosynthesis \cite{polony}.\footnote{The
lifetime should be shorter than $0.1$ sec to
avoid this problem\cite{kawasaki}.}

A solution to the above problem is easily given by introducing the
following superpotential:\footnote{This superpotential is also
discussed in Refs.\cite{nomura},\cite{berkeley}.}
\begin{equation}
  W = h\frac{Z}{M_G} L_{i}HS,
  \label{eq:1}
\end{equation}
where $L_i (i=1-3), H$ and $S$ are supermultiplets of three families of 
lepton doublets, a Higgs
doublet and the sterile neutrino, and $Z$ is a supermultiplet
responsible for SUSY breaking. Then, we get an $A$--term,
\begin{equation}
  {\cal L} = hm_{3/2}\tilde{L_{i}}H\tilde{S},
  \label{eq:2}
\end{equation}
where $\tilde{L_i}$ and $\tilde{S}$ are scalar components of the
supermultiplets $L_i$ and $S$, respectively and $H$ is 
the Higgs boson. We have used that the gravitino mass $m_{3/2}$ is given by the
vacuum-expectation value (vev) of the $F$-component of the supermultiplet
$Z$, that is,
\begin{equation}
  m_{3/2} = \frac{1}{\sqrt{3}M_G}\langle F_{Z}\rangle.
  \label{eq:3}
\end{equation}
Then, the lifetime of the scalar sterile neutrino becomes $\tau \simeq
10^{-26}$ sec and the scalar sterile neutrino is cosmologically
harmless.\footnote{We wish to remark that the operator of Eq. (1)
can provide a possible solution to the moduli problem that is generic
in string theory.  If $S$ is identified as one of the moduli fields
and $L_i$ is $\bar{H}$ in Eq. (1), the cosmological problem associated
with the moduli will be solved, very much in analogy to the scalar neutrino.}
Here, we have assumed $m_{3/2}\simeq 1$ TeV.\footnote{For the large angle
$\nu_e$-$\nu_s$ quasi--vacuum oscillation, it is possible to evade the
cosmological limit by choosing $y_s \sim 3 \times 10^{-13}$, so that
the $\nu_e$-$\nu_s$ mass term is of order $0.1$ eV.  If the direct
$\nu_e$-$\nu_e$ mass term arising 
from the seesaw mechanism is of order $10^{-7}$
eV, the required $\Delta m^2$ for solar neutrinos will be generated.  
Such a scenario is not realized by the
mechanism suggested in this paper as long as all  relevant Yukawa
couplings are O(1).}

We now discuss the SUSY breaking sector. We adopt the SUSY breaking
model found in Ref.\cite{izawa1}, which is based on an SU(2) gauge
theory with four quark doublets, $Q_{\alpha}^i$ ($\alpha =1,2$
and $i=1-4$). We introduce six gauge-singlet supermultiplets $Z_a (a =
1-6)$  and assume the following superpotential:
\begin{equation}
  W = \lambda ^a_{ij}\epsilon_{\alpha\beta}Q_{\alpha}^iQ_{\beta}^jZ_a.
  \label{eq:4}
\end{equation}
It is shown in Ref.\cite{izawa1} that the effective low-energy 
superpotential is given by
\begin{equation}
  W_{eff} = \lambda \Lambda ^2Z.
  \label{eq:5}
\end{equation}
Here, $Z$ is a linear combination of $Z_a$ supermultiplets and $\Lambda$
denotes the dynamical scale of the SU(2) gauge interactions. The Kahler
potential takes, on the other hand,
\begin{equation}
  K= ZZ^* - \frac{k}{2\Lambda ^2}(ZZ^*)^2 + ......,
  \label{eq:6}
\end{equation}
where $k$ is a real constant and the ellipsis denotes higher-order 
terms of $ZZ^*$. If the coupling constant $k$ is positive,
\footnote{This is an important dynamical assumption in this paper. If
the $k$ is negative, the $A$-component of the $Z$ has a vev of order the 
dynamical scale $\Lambda$, which induces too large a Dirac-type mass for 
the sterile neutrino ($m_{\nu D}\sim 1$ keV).} we have a unique vacuum
\begin{equation}
\langle Z \rangle =0,~~~~~\langle F_Z \rangle = \lambda \Lambda ^2.
  \label{eq:7}
\end{equation}
Thus,  SUSY is dynamically broken and the sterile neutrino remains
massless\footnote{We assume the standard Yukawa coupling, $W=fLHS$ to
exactly vanish and consider the case where  $S$ has only grvitationally
suppressed nonrenormalizable interactions. This will be the 
case if $S$ is one of the  moduli fields of string theory. For neutrino
mixing with modulino fields, see e.g. Ref. \cite{smirnov}}. 

A crucial point observed in Ref.\cite{izawa2} is that 
the supergravity effects induce a small shift of the vacuum and 
the $A$-component of the $Z$ has a small nonvanishing vev:
\begin{equation}
\langle Z \rangle \simeq \frac{\Lambda ^2}{\sqrt{3}kM_G} \simeq 
\frac{m_{3/2}}{\lambda k}.
  \label{eq:8}
\end{equation}
Substituting this result, Eq. (\ref{eq:8}), into Eq. (\ref{eq:1}) we obtain
a Dirac-type $\nu_i$-$\nu_s$ mass ($i=e,\mu,\tau$) as
\begin{equation}
 {\cal L} \simeq \frac{h}{\lambda k}\frac{m_{3/2}}{M_G}
\langle H\rangle \nu _i \nu _s + hc.
  \label{eq:9}
\end{equation}
The Dirac-type mass is of order $10^{-4}$ eV for $\frac{\lambda k}{h} 
\sim 1$.\footnote{If one identifies the sterile neutrino with a 
right-handed neutrino and keeps exact lepton-number conservation,
one may have a light Dirac neutrino as discussed in
Ref.\cite{yanagida2}.} 
It is now clear that if the Majorana mass for the active
electron neutrino induced by the seesaw mechanism is of order $10^{-4}$
eV, the present model will naturally reproduce the solar $\nu_e$-$\nu_s$
oscillation.  Since the mechanism suggested here generates
a $\nu_s$-$\nu_i$ ($i= e,\mu,\tau$) mass term, and no direct
$\nu_s$-$\nu_s$ mass term, it turns out that the lighter eigenstate
is predominantly in $\nu_s$, and not in $\nu_e$.  The MSW resonance condition
will not be satisfied for solar neutrinos in this case.  Our scenario
will prefer the quasi--vacuum oscillation solution \cite{concha}
with the inclusion of the Chlorine experiment,
or the energy--independent solutions advocated in Ref. \cite{barger}
excluding the Chlorine experiment.  In either case, 
the other two active neutrinos together with the electron neutrino may 
explain the atmospheric and LSND neutrino oscillations.  While
$\nu_e$-$\nu_s$ MSW resonance does not occur for supernova neutrinos,
$\bar{\nu_e}$-$\bar{\nu_s}$ resonance will occur within the supernova
\cite{arafune}.  However, vacuum  oscillations on its way from
supernova to the Earth will regenerate $\bar{\nu_e}$, but with
its flux reduced by half.\footnote{If $\bar{\nu_s}$ mixes also with 
$\bar{\nu}_{\mu, \tau}$, the supernova $\bar{\nu}_{\mu, \tau}$ are also 
converted into $\bar{\nu_s}$ through the MSW resonances. In this case the
$\bar{\nu_e}$ flux is enhanced by factor 3/2 instead.}
Such a reduction is not inconsistent with
$\bar{\nu_e}$ data from SN1987A, but may be testable with future
supernova neutrinos.  It is interesting to note that the neutrino
data from supernova alone makes the large $\nu_e$-$\nu_s$ mixing
preferable in our scenario, independent of solar neutrino data.

If one supposes a superpotential term $W=(SSZZ/M_G)$, in addition to
Eq. (1), a direct Majorana mass term for the $\nu_s$ of order
$m_{3/2}^2/M_G\sim 10^{-3}$ eV will result.  In this case the small
angle $\nu_e$-$\nu_s$ MSW oscillation may become relevant for solar
neutrinos. However, this operator is less motivated (compared
to the one in Eq. (1)) from the point of view of cosmology.

%%%%%%%%%%%%%%%%%%%%%%%%%%%%%%%%%%%%%%%%%%%%%%%%%%%%%%%%%%%%%%%%%%%%%%
\subsection*{Acknowledgments}
%%%%%%%%%%%%%%%%%%%%%%%%%%%%%%%%%%%%%%%%%%%%%%%%%%%%%%%%%%%%%%%%%%%%%%

One of the authors (T.Y) is grateful to F. Borzumati, H. Murayama and 
Y. Nomura for useful discussions.
K.B acknowledges the Theory Group at Tokyo University for its
warm hospitality.  The  work of T.Y is supported in part by the Grant-in-Aid, 
Priority Area ``Supersymmetry and Unified Theory of Elementary
Particles''(\#707), K.B is supported in part by Department of Energy
Grant No. DE-FG03-98ER41076 and by a grant from the Research Corporation.

\end{document}